\documentclass[aps,prl,showpacs,noshowkeys,amsmath,amssymb,amsfonts,superscriptaddress,twocolumn]{revtex4-2}

\usepackage{amsmath,amssymb,physics,graphicx,bm,siunitx}
\usepackage{xcolor}
\usepackage[colorlinks=true,linkcolor=blue,citecolor=blue,urlcolor=blue]{hyperref}
\usepackage[all]{hypcap} 
\usepackage{graphicx}
\usepackage{amsmath}
\usepackage{bm}
\usepackage{lipsum}
\usepackage{amsmath}
\usepackage{upgreek}
\usepackage{color}
\usepackage{soul} 
\definecolor{darkgreen}{rgb}{0.2,0.6,0.2}

\renewcommand{\vec}[1]{\mathbf{#1}}
\begin{document}
\title{Emergent charge crystallization and frustration in a particle anti-spin Ice}

\author{Renaud Baillou }
\affiliation{Departament de F\'{i}sica de la Mat\`{e}ria Condensada, Universitat de Barcelona, 08028 Spain}
\author{Matthew Terkel}
\affiliation{Departament de F\'{i}sica de la Mat\`{e}ria Condensada, Universitat de Barcelona, 08028 Spain}
\affiliation{Universitat de Barcelona Institute of Complex Systems (UBICS), Universitat de Barcelona, 08028 Spain}
\author{Cristiano Nisoli}
\affiliation{Theoretical Division, Center for NonLinear Studies, Los Alamos National Laboratory, Los Alamos, NM 87545, USA}
\author{Pietro Tierno}
\email{ptierno@ub.edu}
\affiliation{Departament de F\'{i}sica de la Mat\`{e}ria Condensada, Universitat de Barcelona, 08028 Spain}
\affiliation{Universitat de Barcelona Institute of Complex Systems (UBICS), Universitat de Barcelona, 08028 Spain}
\date{\today}
\maketitle

\textbf{Artificial spin ices have transcended their origins in frustrated rare-earth pyrochlores to become a versatile platform for engineering exotic states of matter. Across diverse implementations, from nanomagnets and superconducting vortices to colloids, quantum annealers, liquid crystals, and metamaterials, they are unified by the ice rule, which often leads to degeneracy and constrained disorder by enforcing minimization of the local topological charge. Here, we report the first realization of an “anti-spin ice” in which not only the ice rule does not hold, but its opposite is true as the system seeks to maximize, rather than minimize, spin ice charges. Using fast-rotating, in-plane magnetic fields to generate isotropic attraction between colloidal particles, we invert the conventional paradigm of repulsive interactions in colloidal spin ices. Combining experiments and simulations across standard square and honeycomb lattices as well as novel pentaheptite geometries, we establish rules for order and disorder in the anti-spin ice. With the pentaheptite lattice, we demonstrate that the anti-spin ice system can also exhibit frustration, but of a new kind. This topological charge frustration arises from the lattice connectivity, where networks of unequal, odd-sided polygons suppress charge crystallization at high interaction strength.}

\vspace{2ex}\noindent

Complex systems generally sit at the edge between order and disorder,
where collective behavior emerges from competing interactions. Geometric frustration, the inability to satisfy all local constraints simultaneously, is a basic mechanism that generates such emergent states~\cite{Sadoc1999,Moe06,Hagan2021}. It appears across scales, from water ice~\cite{Bernal1933,Giauque1936} and magnetic materials~\cite{Harris1997,Bramwell2001,Bramwell2009,Jaubert2009,Gardne2010,Giblin2011}, to trapped ions~\cite{Kim2010}, folding proteins~\cite{Bry87,Ferreiro2018}, microgel particles~\cite{Han08,Shokef2011}, and macroscopic robotic gears~\cite{Guo2023}. Frustration often leads to degenerate ground states with residual entropy, suppressed long-range order, and exotic emergent phenomena such as Coulomb phases and topological monopoles~\cite{Castelnovo2008,Fennell2009,Morris2009}.

In this context, artificial spin ice systems (ASIs), namely two-dimensional arrays of interacting magnetic nano-islands~\cite{Wang2006,Nisoli2013,Zhang2013,Skjarvo2020}, were originally introduced as synthetic analogs of frustrated materials inspired by water ice~\cite{Bernal1933,Pauling1935,Giauque1936} and magnetic pyrochlores~\cite{Harris1997,Bramwell2001,Bramwell2009,Jaubert2009,Gardne2010,Giblin2011}. Since then, they have evolved into a versatile platform for realizing exotic collective states, impossible in natural magnets~\cite{Ressouche2009,Rougemaille2011,Zhang2013,gilbert2014emergent,canals2016,Perrin2016,Sklenar2019}. Starting from nanomagnetic systems, the ASI concept has expanded to a variety of other platforms, ranging from macroscopic magnets~\cite{mellado2012macroscopic,teixeira2024macroscopic}, to quantum annealers~\cite{king2021qubit,lopez2023kagome}, patterned superconductors~\cite{latimer2013realization,wang2018switchable}, liquid crystals~\cite{duzgun2021skyrmion}, and even metamechanics~\cite{merrigan2021topologically,sirote2024emergent}.  

So-called ``particle ices"~\cite{ortiz2016,Ortiz2019} originally proposed in~\cite{Libal2006}, are now widely used as an alternative platform for studying geometric frustration that more closely resembles water ice~\cite{Pauling1935}. In these systems, repulsive colloidal particles are gravitationally confined within a lattice of double-well traps aligned along the edges of a graph, meeting at vertices. Due to mutual repulsion, each vertex tries to expel nearby particles, but this cannot be achieved globally. The ice rule thus emerges as a collective compromise, with each vertex receiving a balanced number of colloids. In a square geometry, for example, each vertex hosts two colloids close to and two far from the vertex center, \textit{exactly as in water ice}. 
Previous theoretical arguments showed that, for lattices with single coordination number $z$ and central symmetry, particle-based ices exhibit an ice-like manifold~\cite{Nisoli2014,nisoli2018unexpected}. 
Moreover, unlike ASIs, colloidal ices allow \textit{in-situ} tuning of interparticle interactions via an external field, avoiding the need to fabricate new lithographic structures with varying trap spacing.

\begin{figure*}[t]
\includegraphics[width=0.85\textwidth]{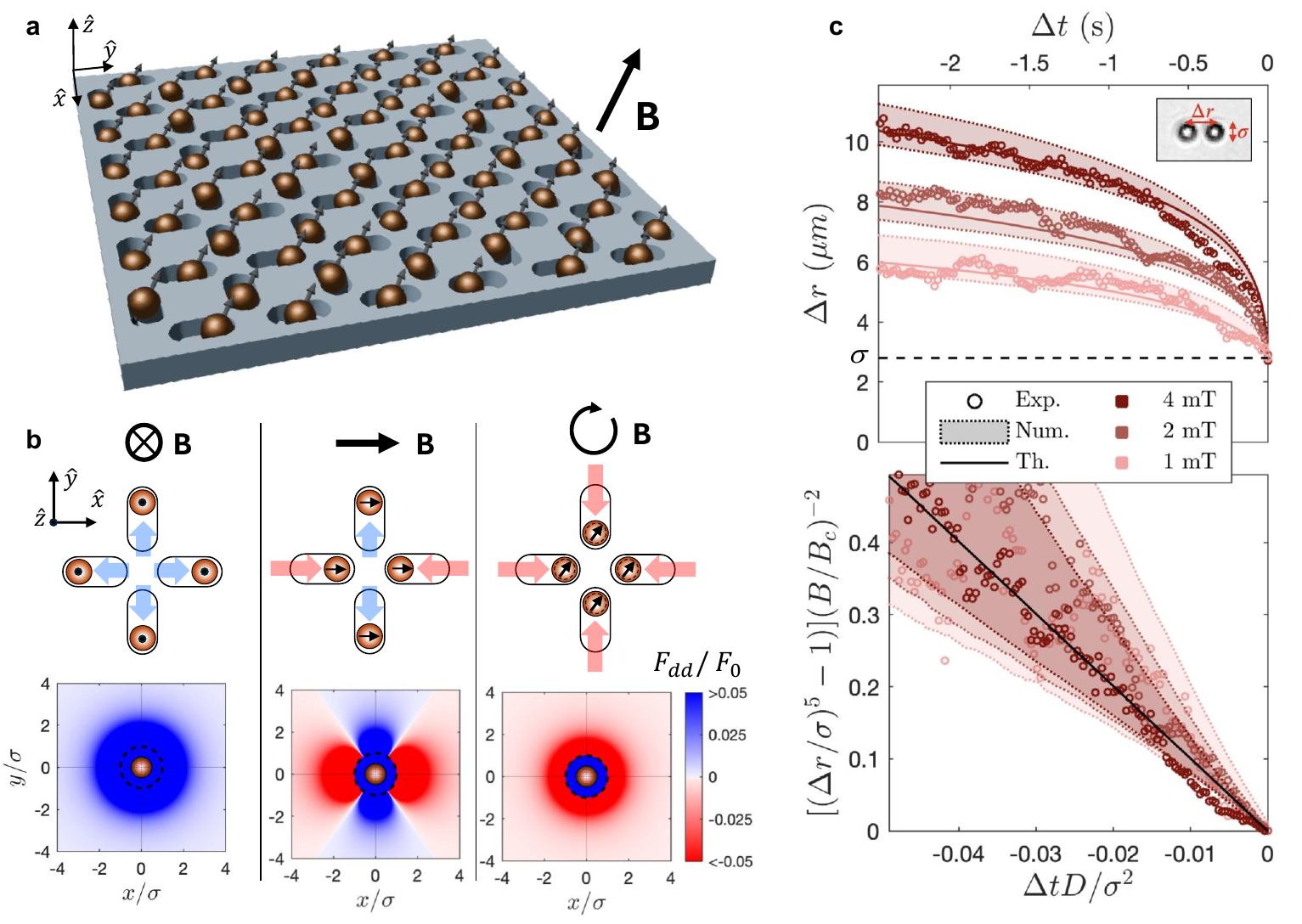}
\caption{\textbf{Repulsive and attractive interactions in the particle ice.}  (\textbf{a}) Schematic showing an ensemble of paramagnetic colloidal particles confined by gravity within a square lattice of cylindrical wells and subjected to an external magnetic field $\bm{B}$. The field induces an equal dipole moment $\bm{m}$ in each particle. (\textbf{b}) Top: Equilibrium states in a vertex of a square lattice for colloids subjected to an out-of-plane (left), an in-plane (middle) and a rotating (right) magnetic field.
Bottom: Magnetic dipolar force $F_{dd}$ acting on a particle, with repulsion
(attraction) highlighted in blue (red). Here  $F_0= 0.26$ pN.  (\textbf{c}) Measurements of the attractive interactions between two particles under a rotating field with frequency $f=50$Hz and different field amplitudes. Top: Separation distance $\Delta r$ versus time interval $\Delta t=t-t_{\sigma}$, with $t_{\sigma}= t(\Delta r = \sigma)$ and  $\sigma=2.8 \rm{\mu m}$. Inset shows microscope images of two particles, see Supplementary Video 1. 
Bottom: re-scaled separation,
being $D=0.156 \, \rm{\mu m^2 s^{-1}}$ the particle diffusion coefficient and $B_{c}=12\sqrt{2\eta D\mu_0 }/ (\chi \sigma)=0.106$ mT the field amplitude required to keep a pair of approaching particles in close contact.
Here calculate $B_{c}$ by comparing the typical Brownian speed $v_B=D/\sigma$ to the speed of approach of a magnetic particle $v$ due to the dipolar repulsion with a neighbor at $\Delta r=\sigma$. In both graphs, scattered symbols are experimental data, shaded regions indicate numerical simulations results with their standard deviation.}
\label{figure1}
\end{figure*}

Since their introduction, particle ice systems have been realized using isotropic repulsive colloids,  either due to electrostatic~\cite{Libal2006,Libal2012} or magnetic~\cite{ortiz2016,Loehr2016} interactions. Although an early theoretical proposal considered attractive particles on disordered graphs in the context of wealth distribution in social science~\cite{mahault2017emergent}, no study to date has explored attractive colloids on a regular lattice in a physical system. 

We demonstrate here the first ``anti-spin ice", made of  trapped colloids which experience tunable isotropic {\em attractive} interactions. We show that the switch of the interaction sign does not simply produce the reverse situation of the repulsive case, but it allows the exploration of a new set of  phenomenology related to  crystallization, and 
frustration of topological charges. We show that in simple bipartite crystals, the low energy states of the attractive spin ice emerge from a spontaneous symmetry breaking leading to staggered organization of topological charges, in which half of the vertices accumulate all the colloids.  However, in a non-bipartite geometries with different odd-sided polygons, such as
the non-Archimedean  pentaheptite lattice,  we show that a new form of frustration emerges. The crystallization of topological charges is frustrated, impeding charge crystallization and leading to a degenerate manifold.   

\section*{Attractive interactions in the particle ice.}
The main features of a colloidal ice are shown in Fig.~\ref{figure1}(a)
: interacting paramagnetic colloids are confined by gravity in topographic semi-cylindrical wells, each filled with a single particle.
An external magnetic field $\bm{B}$ induces
equal dipoles in each particle, $\bm{m}=V\chi_V \bm{B}/\mu_0$,
being $V= (\pi \sigma^3) /6$ the particle volume, $\sigma$ its diameter, $\chi_V$ the magnetic volume susceptibility and $\mu_0$ the permeability of the medium.
Thus, the interaction between two magnetic moments $\bm{m}_{i,j}$ located at a separation distance $r=|\bm{r}_i-\bm{r}_j|$ within the same plane is given by,
\begin{equation}
U_{dd}=\frac{\mu_0}{4\pi} \left[ \frac{\bm{m}_i \cdot \bm{m}_j}{r^3}- \frac{ 3(\bm{m}_i \cdot \bm{r})(\bm{m}_j \cdot \bm{r})}{r^5} \right] 
\label{interactions}
\end{equation}
which is 
attractive (repulsive) for particles with moments parallel (normal) to $\bm{r}$. 
In a repulsive particle ice, the magnetic field is
applied perpendicular to the sample pane, $\bm{B}=B\hat{\bm{z}}$ 
and Eq.~\ref{interactions} reduces to an isotropic, repulsive potential, $U_{dd}=\mu_0 m^2/(4\pi r^3)$~\cite{ortiz2016,Loehr2016}.
However, as illustrated by the sequence of images in Fig.~\ref{figure1}(b), the induced dipolar interactions between the particles can be tuned not only in magnitude but also in sign by varying the orientation of $\bm{B}$. For instance, when an in-plane field $\bm{B}=B\hat{\bm{x}}$ is applied the interactions become anisotropic at the vertex, and thus, colloids with
head-to-tail magnetic moments attract, while those with moments side by side repel, middle of Fig.~\ref{figure1}(b).

\begin{figure*}[t]
\includegraphics[width =\textwidth]{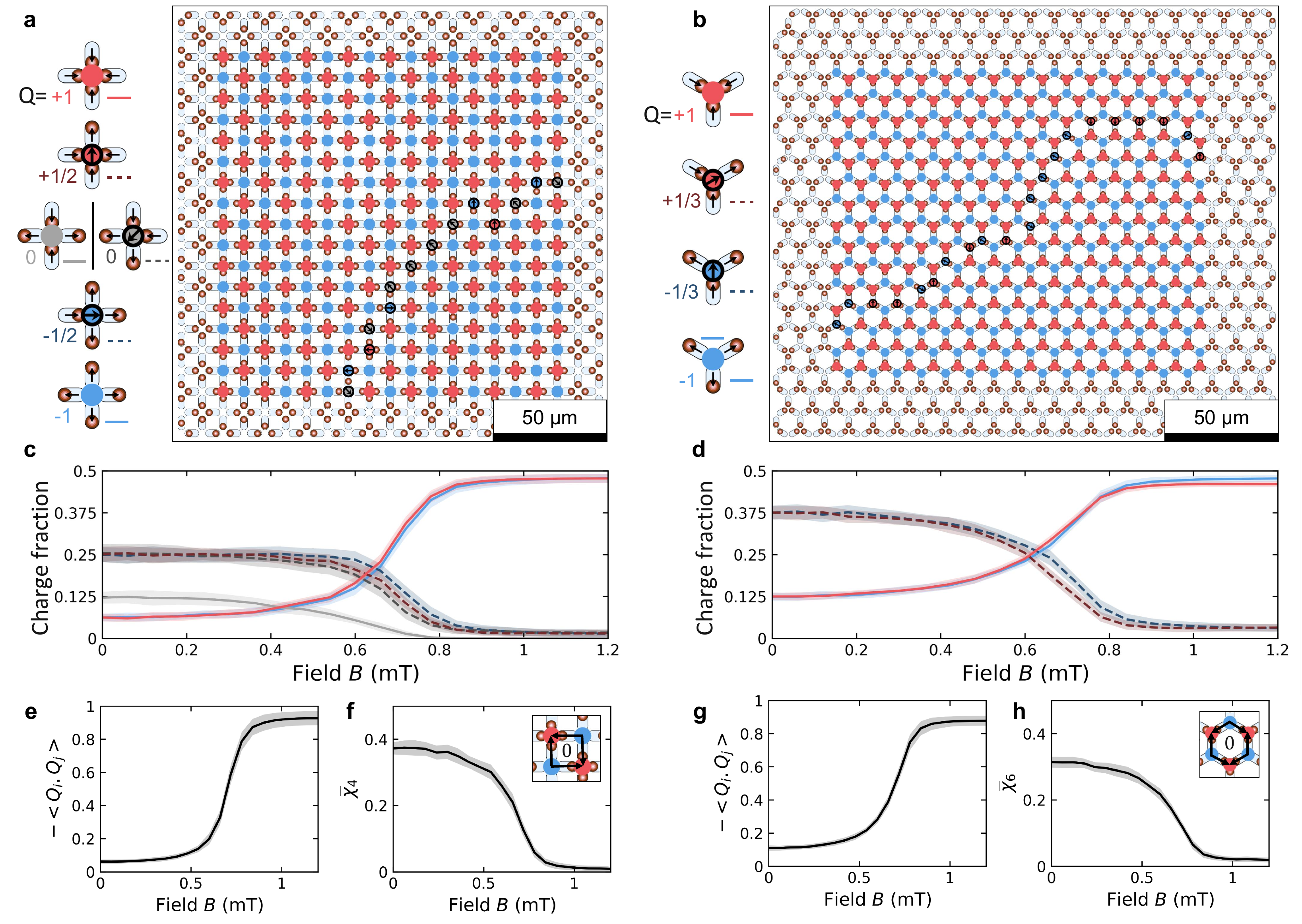}
\caption{\textbf{Charge crystals in bipartite lattices: Numerical results.}  (\textbf{a},\textbf{b}) Snapshots of the low energy states of the square, $z=4$ (Supplementary Video 2)  and the honeycomb, $z=3$ (Supplementary Video 3)  particle anti-spin ice ($f=50$ Hz, $\alpha = 1.2\cdot 10^{-3} \, \rm{mT \, s^{-1}}$, $B_{\rm{max}}=1.2$ mT). Legends on the side show the normalized topological charge $Q=q/z$ for each vertex.  
Ice rule vertices have minimal absolute charge which is $Q=0$ ($Q=\pm 1/3$) for the square (honeycomb), but the ground state in the anti-spin ice features crystallization of high charges ( $Q=\pm1$) for both lattices.  These charges are highlighted in the central region of the two images. The vertices 
with $Q=\pm 1/2$ and one of the two $Q=0$ vertex type in the square 
as the  $Q=\pm 1/3$ for the honeycomb present a net moment shown by a small arrow.
(\textbf{c},\textbf{d}) Fraction of topological charges for the square (c) and the honeycomb (d) lattices versus rotating field amplitude $B$. For both cases, charge crystallization at high field is evidenced by the similar fraction of $Q=\pm 1$ defects (continuous lines) and disappearance of the low charged vertices   (dashed lines).
(\textbf{e},\textbf{g}) Charge-charge correlation $\langle Q_i\cdot Q_j \rangle$ and (\textbf{f},\textbf{h}) mean chirality $\bar{\chi}$ for the square (e,f) and the honeycomb (g,h) anti-spin ice. 
Small insets in (f,h) show the spins associated to each particle  
in a plaquette which gives $\bar{\chi}_{4,6}=0$.}
\label{figure2}
\end{figure*}

\begin{figure*}[t]
\includegraphics[width=\textwidth]{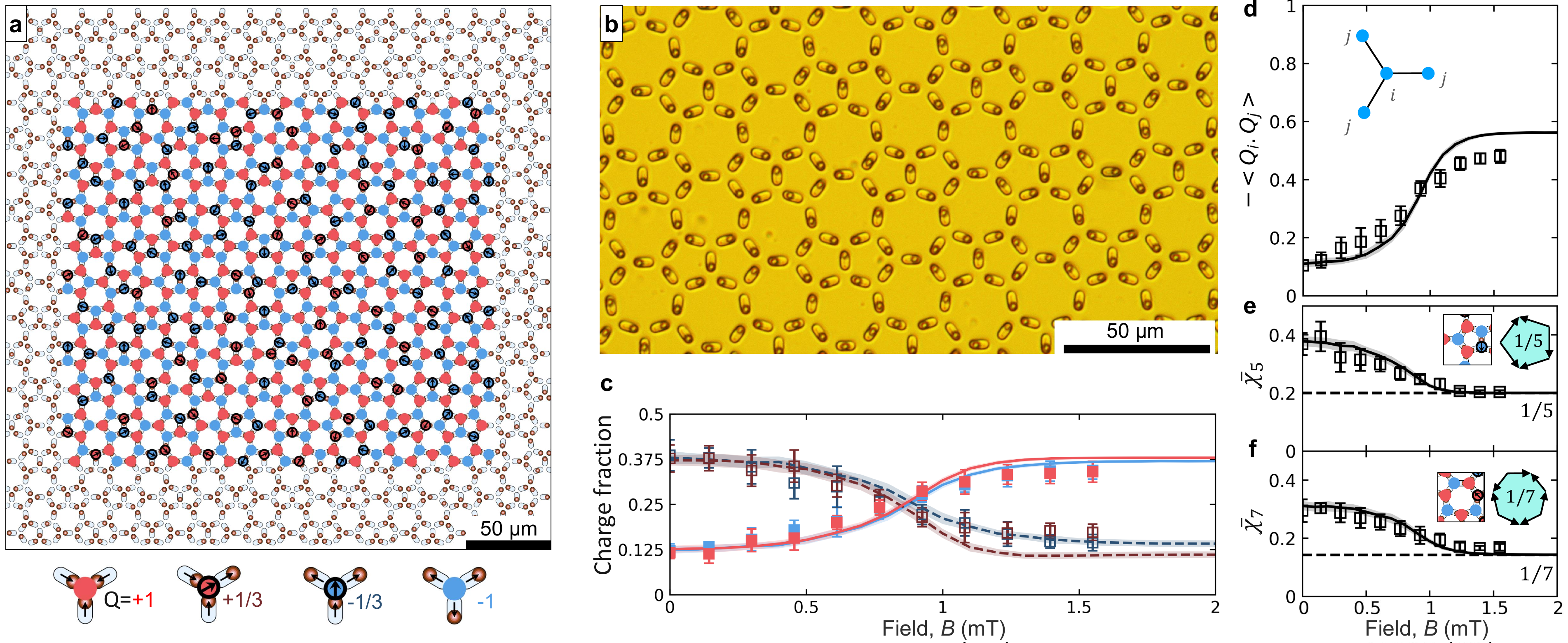}
\caption{\textbf{Frustrated pentaheptite anti-spin ice.} 
\textbf{a} 
Simulation snapshot of the pentaheptite anti-spin ice ($f=50$ Hz, $\alpha = 2\cdot 10^{-3} \, \rm{mT \, s^{-1}}$, $B_{\rm{max}}=2$ mT). Overlaid to the particle positions are topological charges classified following the legend at the bottom (Supplementary Video 4).   
\textbf{b} Experimental realization of the Pentaheptite anti-spin ice using paramagnetic colloidal particles trapped within lithographic elliptical wells. 
Supplementary Video 5 shows the lattice subjected to a rotating magnetic field 
($f=1$ Hz, $\alpha = 6.25\cdot 10^{-3} \, \rm{mT \, s^{-1}}$, $B_{\rm{max}}=3.75$ mT).
\textbf{c} Vertex fraction versus rotating field amplitude $B$ showing the increase of  $Q=\pm 1$  high topological charges and the decrease of 
ice rules $Q=\pm 1/3$ which did not vanishes at large amplitudes due to the topological frustration.
\textbf{d-e} Charge-charge correlation $-\langle Q_i\cdot Q_j \rangle$ (d)
and mean chirality (e) of pentagonal $\chi_5$ (top) and heptagonal $\chi_7$ (bottom) plaquettes
versus rotating field amplitude $B$ ($f=50$ Hz) for the pentaheptite anti-spin ice. 
The dashed lines in (e) represents the predicted value 
for the lowest energy states.
In (c-e) symbols are experimental data,  and continuous line result from numerical simulations. The good agreement was obtained by dividing the experimental values of the field to $2.4$ (see text).}
\label{figure3}
\end{figure*}

We induce isotropic attractive interactions by applying a rapidly spinning, in-plane 
rotating magnetic field 
circularly polarized in the ($\hat{\bm{x}},\hat{\bm{y}}$) plane,
\begin{equation}
\bm{B} \equiv B[\cos{(2\pi f t)}\hat{\bm{x}}-\sin{(2\pi f t)}\hat{\bm{y}}]
\label{field}
\end{equation}
being $f$ the driving frequency  and $B$ the field amplitude. 
Indeed, by averaging  Eq.~\ref{interactions} over a field period, one can obtain an isotropic attraction, $\langle U_{dd} \rangle=-\mu_0 m^2/(8\pi r^3)$ which induces clustering 
of particles, as previously shown with magnetic colloids dispersed in water~\cite{Helgesen1990,Tierno2007,Du2013,Yan2015}.
To effectively verify this attraction, we performed experiments 
by monitoring the collapse of a pair of unconfined particles  ($\sigma=2.8 \rm{\mu m}$)
subjected to a rotating field with $f=50$Hz and at different field amplitudes, Fig.~\ref{figure1}(c) and Supplementary Video 1.
In the overdamped limit, the isotropic attraction can be balanced with 
the relative frictional force in water, $\bm{F}_{dd}+\bm{F}_v =\bm{0}$,
where the velocity of approach of each particle is given by 
$\bm{u}=2\bm{F}_v/\gamma$, being $\gamma = 3 \pi \eta \sigma$ the friction coefficient and $\eta=10^{-3} \, \rm{Pa \cdot s}$
the viscosity of the medium (water). From this balance equation, 
the particle separation distance evolves as
$\Delta r = \left[ \Delta r_0 - 5 \alpha  (t-t_0)  \right]^{1/5}$ where $\alpha=3\mu_0 m^2/(16 \pi \gamma)$ and $t_0$ is the initial time.
The continuous lines in Fig.~\ref{figure1}(c) effectively show that, by rescaling all measurements, 
the relative distance between the pair algebraically decreases with time until it is close to contact at $t_{\sigma}=t(\Delta r = \sigma)$. 
Having found experimentally a way to introduce isotropic attractive forces at the pair interaction level, we then implement them directly in a particle ice.  We show that these interactions profoundly affect the low-energy states of different geometries, giving rise to unexpected phenomena different from those observed for the repulsive case. 

\section*{Topological charges in bipartite lattices.}

We start considering the effect of attractive interactions in simple bipartite lattices, i.e. lattices which can be decomposed into two distinct, compenetrating sublattices.  The most investigated bipartite geometries with repulsive colloids~\cite{Libal2017,Libal2018,Erdal2020,Ostinato2025} and ASIs~\cite{Moller2006,Nisoli2010,Budrikis2011,Rougemaille2011,Budrikis2012,Levis2013,Perrin2016,canals2016} are the  
square, coordination number $z=4$, and the honeycomb, $z=3$. 
For the first, the two sublattices are similar (both $z=4$),
while the honeycomb can be decomposed as the union of two displaced triangular lattices ($z=6$). 

In a colloidal ice the binary nature of the particle position within a 
cylindrical trap allows to assign a pseudo spin variable, $\sigma_i=\pm 1$
which is $1$ ($-1$) when the particle points toward (away from) the vertex center. This maps the colloidal system onto an Ising spin model,
and allows us to define a topological charge to each vertex as $q=2n-z$
being $n$ the number of spins that point toward the vertex center.  Using this mapping, one can identify $6$ and $4$ vertex types
to the square and the honeycomb lattices, respectively, Fig.~\ref{figure2}.
In analogy to water ice~\cite{Bernal1933}, the low energy states are expected to fulfill the ``ice-rules'' which corresponds to the minimization of $|q|$. Topological charges are assigned to vertices that violate such rule, featuring $|q|>0$ .  These defects follow the conservation of the topological charge and can only be created or destroyed in pairs~\cite{Loehr2016,Libal20182}. 

For both the repulsive colloid and the ASI systems, 
the low energy states of the square and the honeycomb lattices 
tend to follow the ice rules, reducing as much as possible the number of  topological charges. 
In contrast to that, we find a completely different situation in our particle anti-spin ice, where highly charged vertices are preferred and crystallize at large interaction strengths. This situation is shown in Fig.~\ref{figure2}(a) for the square (Supplementary Video 2) and Fig.~\ref{figure2}(b) the honeycomb systems (Supplementary Video 3). 
In both cases, we randomly initialize the particle positions in each topographic well and induce attractive magnetic interactions using
the rotating field by slowly increasing its amplitude to a maximum value $B_{\rm{max}}=1.2$ mT at a rate of $\alpha= 1.2\cdot 10^{-4} \, \rm{mT \, s^{-1}}$. 
To compare both lattices, we define a normalized topological charge: $Q = q/z$, 
and show the corresponding values in the central legend in Fig.~\ref{figure2}. In the attractive case, both types of lattices display a unique ground state (GS) 
characterized by an alternating crystallization of $Q= \pm 1$ charges. Few low-energy vertices ($Q=\pm 1/2$ and biased $Q=0$ for the square, $Q=\pm 1/3$ for the honeycomb) are trapped along grain boundaries that could eventually be eliminated in the limit of adiabatically slow driving, $\alpha \to 0$.

In conventional square ice, the low-energy vertices obey the two-in/two-out rule, with two spins pointing toward the vertex center and two pointing away. Depending on the specific spin arrangement, two types of $Q=0$ vertices can be distinguished: unbiased vertices, which lead to an antiferromagnetic ground state, and biased vertices, which produce a ferromagnetic ground state. The biased vertices carry a net magnetic moment, as indicated by the central arrow in the legend of Fig.~\ref{figure2}(a).
For square anti-spin ice, increasing the field amplitude rapidly suppresses the unbiased $Q=0$ vertices, which vanish at $B = 0.8$ mT. Above $B = 1$ mT, the biased $Q=0$ vertices (dashed gray line) and the $Q = \pm 1/2$ vertices also disappear. Thus, the attractive square system exhibits the opposite behaviour as in the repulsive case~\cite{ortiz2016}, in which the population of highly charged vertices is minimized rather than maximized.

However, a fundamentally different picture emerges for the honeycomb ice, shown in Figs.~\ref{figure2}(b,d). In the repulsive colloidal ice~\cite{Libal20182,Cunuder2019} as in ASI~\cite{Qi2008,Ladak2010,Mengotti2011,Zhang2013}, this geometry does not possess a unique ground state (GS) that satisfies all nearest-neighbor interactions, but each vertex of the three islands will have a net magnetic charge, $Q=\pm 1/3$. In contrast, here we find that the switch in the interaction sign induces again an ordered 
GS with charge crystallization. Thus, for attractive interactions, the particle location at each vertex is governed more by the type of lattice topology rather than by its energetics.
The emergence of long-range crystallization is further supported by the charge-charge correlations shown in Figs.~\ref{figure2}(e,g) for both lattices. The negative values of $\langle Q_i Q_j \rangle$ indicate that neighboring vertices preferentially host opposite topological charges, revealing an effective tendency towards staggered charge ordering. At large field amplitudes, the system becomes fully anticorrelated, with $\langle Q_i \cdot Q_j \rangle \to -1$.
This behavior is also captured by the net chirality, $\chi_i$, associated to each elementary plaquette $i$ of the lattice.
Using the location of the spins associated with each particle, we measure the average absolute chirality, $\bar{\chi} = \frac{1}{N_{pl}} \sum_i |\chi_i|$, where $N_{pl}$ is the number of plaquettes. 
In the repulsive colloidal ice, particles at the vertices satisfy the ice rule and develop an emergent chirality 
which leads to its maximum value, $\bar{\chi} = 1$ for the ground state of the square ice~\cite{ortiz2016,Loehr2016}, and to 
$\bar{\chi} =2/3$ for the 
spin solid phase in the honeycomb lattice~\cite{Libal20182,Cunuder2019}. 
Here, by contrast, the  charge crystallization suppresses chirality, driving $\chi \to 0$ at large interaction strengths, as shown for both the square, Fig.~\ref{figure2}(f),  and the honeycomb, Fig.~\ref{figure2}(h), anti-spin ice.

\section*{The frustrated pentaheptite lattice.}
With attractive interactions, bipartite geometries exhibit a unique GS characterized by charge crystallization. This naturally raises the question of whether geometric frustration is entirely absent in the anti-spin ice, or if some lattice geometries can still host a degenerate GS. To address this question, we first examine non-bipartite lattices with even coordination numbers.  The simplest are the triangular ($z=6$) and kagome ($z=4$) lattices, which are tripartite: their sites can be divided into three sublattices such that nearest neighbours always belong to different sublattices. However, we find that in the presence of attractive interactions both of them still exhibit a unique GS, as shown in Figs.~\ref{figureSI1}(c,d) in the Method Section. Yet, we find that degeneracy can be recovered in the anti-spin ice by considering a non-bipartite geometry with odd plaquettes of unequal sizes.
Figure~\ref{figure3} shows the pentaheptite lattice, a 
$z=3$ non-Archimedean geometry obtained 
by tiling the plane with a repeating arrangement of six heptagons (7-sided) and two pentagons (5-sided)~\cite{Deza2000}. This lattice can be found in a variety of condensed-matter settings, including graphene~\cite{Crespi1996,Lin2014}, nanotubes~\cite{Nardelli1998}, nanoribbons~\cite{Fan2019}, and materials exhibiting chiral spin-liquid behavior~\cite{Peri2020}, although it was not previously employed in spin ice systems. Among the different ways to construct the pentaheptite lattice, we adopt the one characterized by an equal distance between neighboring vertices (see details in Methods Section), in order to preserve the same separation between adjacent attractive particles. 
With this choice, we can approximate the number of vertex types to four, analogously to the honeycomb case, at the bottom of Fig.~\ref{figure3}(a).

Typical low-energy states obtained after slowly annealing the system are shown in Fig.~\ref{figure3}(a) for simulations (Supplementary Video 4) and Fig.~\ref{figure3}(b) for experiments (Supplementary Video 5). In both cases, we construct the lattice by placing randomly the particles in each cylindrical trap, resulting in a low fraction of 
$Q= \pm 1$   vertices ($\phi=0.125$) and a larger fraction of 
$Q= \pm 1/3$ ($\phi=0.375$). 
Fig.~\ref{figure3}(c) shows the corresponding evolution of the 
fraction of topological charges $\phi$ as attraction increases by increeasing the amplitude of the rotating field. 
For $B> 0.9$ mT, the charge populations invert: high charge vertices $Q= \pm 1$ proliferate at the expense of the ice-rule ones, $Q= \pm 1/3$. However, unlike the bipartite geometries, the topological charges do not vanish completely, but persist in the form of one or two defects per plaquette, reaching a lower fraction $\phi\sim 0.12$, Fig.~\ref{figure3}(a). The odd number of 
bonds in each plaquette frustrates any perfectly staggered arrangement of the
$Q= \pm 1$ and the system cannot globally realize the local energetic tendency of the anti-spin ice to maximize the vertex charge. 
The final fraction of vertices can be understood by considering that the  basic unit cell of the penthaheptite is made of $2$ pentagons and $2$ heptagons, and
will have a total of $8$ vertices, see Fig.~\ref{figureSI2} in the Method Section.
With attractive interactions, the system maximizes the vertex charges, 
reducing to a minimum of $2$ ice rules vertices per unit cell. 
As a result, the charge fractions appear as multiples of $1/8$,
giving $\phi=1/8=0.125$ for the $Q= \pm 1/3$ and $\phi=6/8$
for the high charged ones, $Q= \pm 1$.
This leads to fixed vertex fractions imposed by topology rather than thermal statistics.

The absence of long-range charge crystallization is further evidenced by the correlation shown in Fig.~\ref{figure3}(d). Similarly to bipartite geometries, the pentaheptite lattice also displays an effective tendency towards staggered charge ordering, although $\langle  Q_i\cdot Q_j \rangle$ never reduces to $-1$. Indeed, the topological frustration in the pentaheptite anti-spin ice is reflected in short-range charge ordering. Geometric frustration emerges from the presence of an uneven number of bonds per plaquette which prevents perfect global antiferro-like charge ordering. Again, the resulting correlation values at high field may be directly explained from the basic crystallographic arrangement of the pentaheptite unit cells, Fig.~\ref{figureSI2} in the Method Section.
This effect is further captured by the mean plaquette chirality, $\bar{\chi}$.
Since the pentaheptite lattice contains pentagonal and heptagonal plaquettes, Fig.~\ref{figure3}(e) reports  $\bar{\chi}_5$ and $\bar{\chi}_7$ separately.  In both cases, the chirality decreases at large field amplitudes as a consequence of charge crystallization. However, it does not vanish completely, since each odd plaquette hosts
at least one of the ice-rule vertices ($Q=\pm 1/3$) in the low-energy manifold. Thus, these vertices are located on one of the five sites of a pentagon ($1/5$) or one of the seven sites of a heptagon ($1/7$). 
We finally note that good agreement between experiments (scattered data)
and simulations (continuous lines) in Figs.~\ref{figure3}(c-f) was obtained by rescaling the experimental field values as $B_{\rm{exp}}=2.4 B_{\rm{sim}}$. This discrepancy arises from the fact that the shortest distance between the particles at a vertex in the fabricated penthaeptite lattice, $d_{\rm{exp}}=5.9 \, \rm{\mu m}$, was found to be greater than the simulated one, $d_{\rm{sim}}=4.6 \, \rm{\mu m}$. This $d_{\rm{exp}}$ requires a higher field amplitude to reach the same attractive forces between the particles, which scale as $F_{dd}\sim B^2 \chi_V^2/r^4$. In addition, considering that the magnetic volume susceptibility of the particles $\chi_V$ is a material property that can change from one batch to the other, the factor $B_{\rm{exp}}/B_{\rm{sim}}=2.4 $ results by assuming that $\chi^{\rm{exp}}_V\sim 0.54$, similar to previous works~\cite{ortiz2016,Libal2018,Erdal2020}.

\section*{Conclusions}

We have realized the first experimental demonstration of an anti-spin ice, in which isotropically attractive colloidal interactions invert the conventional ice-rule paradigm. In bipartite lattices, this inversion produces staggered crystallization of maximal topological charges. On the non-bipartite pentaheptite lattice, however, the same mechanism gives rise to a novel form of frustration—topological charge frustration—preventing global charge ordering and stabilizing a degenerate manifold.
In this sense, the pentaheptite anti-spin ice provides a minimal realization of a system where frustration originates from the impossibility of satisfying a local maximization principle across odd plaquettes. Unlike conventional geometric frustration, which typically suppresses high charge states, here frustration arises precisely because the system attempts to crystallize them.

These findings broaden the conceptual framework for artificial ice systems. Rather than viewing frustration as intrinsically tied to minimizing local constraints, our work shows that it can also emerge when maximizing local topological quantities, provided the lattice forbids their global arrangement. This dual perspective suggests that frustration is fundamentally a property of compatibility between local energetic rules and global topology.

By combining tunable colloidal interactions with non-bipartite lattice geometries, our work establishes that frustration in particle ice can be engineered not only by lattice geometry but also by tuning the sign of the pair interactions. More broadly, 
the anti-spin ice opens a route toward programmable classical topological phases. This platform may enable controlled exploration of constrained dynamics, emergent gauge descriptions, and charge transport phenomena in systems where frustration is not an accident of materials chemistry, but a design principle.

\section*{References}
%


\newpage
\clearpage

\noindent\textbf{\large Methods}\\

\noindent\textbf{Fabrication of the lithographic structures.}
The different lattices used in this work were fabricated using soft lithography following previous works~\cite{Gallo20233}.
In particular, we design a $5''$ chrome photomask ($4500$ kdpi, JD Photodata) with the desired pattern and then transfer it to a $10$ cm silica wafer. 
Before transferring the mask, the wafer was dehydrated at $200\,^{\circ}$C for $15$ min and plasma-cleaned (Harrick, PCD-002-CE). SU-8 3005 photoresist ($\sim 5 \rm{\mu m}$ thickness) was spin-coated (Laurell Tech, WS-650) on the wafer  and soft-baked at $95\,^{\circ}$C for $2$ min. After cooling to room temperature, the wafer and mask were aligned in a mask aligner (SÜSS Microtec MJB4, i-line with wavelength $\lambda=365$ nm ) and exposed to UV light for $5.4$s at a power of $100 \rm{mJ \, cm^{-2}}$. Post-exposure baking was performed at $65\,^{\circ}$C for $1$min and $95\,^{\circ}$C for $1$ min, followed by development in propylene glycol monomethyl ether acetate for $30$ s, rinsing with isopropanol, and dried with N$_2$ . The wafer was finally hard-baked at $95\,^{\circ}$C for $20$ min and $65\,^{\circ}$C for $10$ min.

For Polydimethylsiloxane  (PDMS) replication, the wafer was plasma-treated and silanized under vacuum with trichloro(1H,1H,2H,2H-perfluorooctyl)silane for $1$ h. PDMS (SYLGARD 184, Sigma-Aldrich) was prepared in a $10:1$ base-to-curing agent ratio, mixed, degassed, and spin-coated ($\sim 25 \, \rm{\mu m}$) onto the wafer. After leveling overnight, the PDMS was cured at $65\,^{\circ}$C for $4$ h in a vacuum oven (Memmert V029). The glass coverslips (Menzel-Gläser) were cleaned, plasma-treated, and placed face-down on the microstructured PDMS, forming a bonded thin membrane. The microstructured membrane was carefully peeled from the wafer and enclosed within a $15 \times 15 \times 0.26$ mm plastic spacer (Gene Frame).

\noindent\textbf{Experimental setup.}  
We use monodisperse 
paramagnetic polystyrene particles with diameter $\sigma=2.8\,\rm{\mu m}$  (Dynabeads,
M270).  The particles were dispersed in highly deionized
water (Milli-Q water) at room temperature $T=293$ K, and the suspension is
confined within a fluidic cell assembled with two coverslips separated
by $\sim 260 \, \rm{\mu m}$.  The cell was placed on the stage of a upright  
light microscope (Eclipse Ni, Nikon) equipped with different magnification objectives ($100\times$, $40\times$) and a $0.45\times$ TV
adapter.  To apply a rotating field in a plane, the two pair of coils were connected
with a wave generator (TGA1244, TTi) feeding a power amplifier (AMP-1800, AKIYAMA or BOP 20-10M, Kepco)
and two sinusoidal currents with $\pi/2$ radians phase-shift were passed through the coils.
The particle dynamics was recorded using a CCD camera (Basler Scout scA640-74fc) working at a maximum frame
per second (fps) of $50$ fps for color images and $75$ fps for black and white. The positions of the particles were obtained from the analysis of .AVI videos analyzed by using a custom Matlab program.

\vspace{5px}
\noindent\textbf{Numerical simulations.}  
We perform Brownian dynamic simulations
using experimental parameters as inputs. In particular, we model 
$N$ paramagnetic colloidal particles of diameter $\sigma=2.8 \rm{\mu m}$, 
magnetic volume susceptibility $\chi_V=0.8$ and 
confined within a lattice of semi-cylindrical wells at a one-to-one filling,
Fig.~\ref{figure1}. 
For each particle at position $\bm{r}_i$, $i=1...N$, 
we integrate  the overdamped equations of motion:
\begin{equation}
\gamma d\bm{r}_i/dt  = \bm{F}_i^{\mathrm{dd}} + \bm{F}_i^{\mathrm{t}} +   \bm{\xi}
\label{eqmotion}
\end{equation}
being $\gamma = 0.032 \mathrm{pN\ s}\ \mu{}\mathrm{m}^{-1}$ the friction coefficient. The magnetic dipolar force $\bm{F}_i^{\mathrm{dd}}$  is calculated from the orientation of the magnetic moments, 
$\bm{F}_i^{\mathrm{dd}}=-\sum_{j\neq i} \partial U_{dd}(\bm{m},\bm{r}_{ij})/\partial \bm{r}$, where $U_{dd}$ is given by Eq.~\ref{interactions}. This force  is induced by an external, in-plane rotating magnetic field with frequency $f=50$ Hz and increasing amplitude $B$ obtained from a slow ramp $\alpha$.
Dipolar interactions are calculated in an iterative way to take into account of the mutual induction between the particles.
Long-range effects are considered by using a large cutoff distance of 200 $\, \rm{\mu m}$.

The force from the semi-cylindrical confinement  
$\bm{F}_i^{\mathrm{t}}$ that acts on a colloid $i$
is modeled as: 
\begin{align}
\vec{F}_i^{\mathrm{t}} =& -\hat{\mathbf{e}}_{\perp} k r_\perp + \nonumber\\
&\hat{\mathbf{e}}_{\parallel}
	\begin{cases}
		0
			& \left|r_\parallel\right| \leq \frac{\lambda}{2} \\
		k \left(\frac{\lambda}{2}-\left|r_\parallel\right|\right) \mathrm{sign} \left(r_\parallel\right)
			& \left|r_\parallel\right| > \frac{\lambda}{2}
	\end{cases}
	\label{pot1}
\end{align}
where $r_\parallel$ and $r_\perp$ are components of a vector $\vec{r}$ parallel and perpendicular with respect to the line of length $\lambda$ that joins the two minima in the double well. The stiffness $k = 6 \cdot 10^{-3} \ \rm{ pN/nm}$ keeps the particle confined to the elongated region around the center of the trap. 

Finally, $\bm{\xi}$ represents a random force due to thermal fluctuations, with zero mean, $\left< \bm{\eta} \right> = 0$ and delta correlated, 
$\left< \bm{\eta} \left(t\right) \bm{\eta} \left(t'\right)\right> = 2k_\mathrm{B}T\gamma \delta(t-t')$, being $T=300\mathrm{K}$ the thermodynamic temperature. 

We consider three types of lattices, the square ($N=840$, lattice constant $a=3\, \sigma$), the honeycomb ($N=1269$, $a=4.5\, \sigma$)
and the pentaheptite ($N=1918$, $D_x=11.99 \, \sigma$ see later for  definition of $D_x$). 
In all of them we use elliptical wells of length $2\sigma$ but with no central hill 
contrary to previous works~\cite{ortiz2016,Loehr2016}.
Indeed, in the case of the repulsive colloidal ice, the binary nature of particle positions was enforced by using a double-well potential with a central barrier, which prevents repulsive particles from localizing at the center of the wells at large field amplitudes~\cite{Gallo20212}. This barrier is no longer required when the interactions become attractive, as the central position becomes unstable rather than stable.

The simulations are performed using a constant time step
of $1$ ms, and by applying a rotating magnetic field with fixed frequency $f=50$Hz and increasing amplitude $B$  at a slow rate $\alpha = 2. \cdot 10^{-4} \rm{mT s^{-1}}$.

\begin{figure*}[t]
\includegraphics[width=\textwidth]{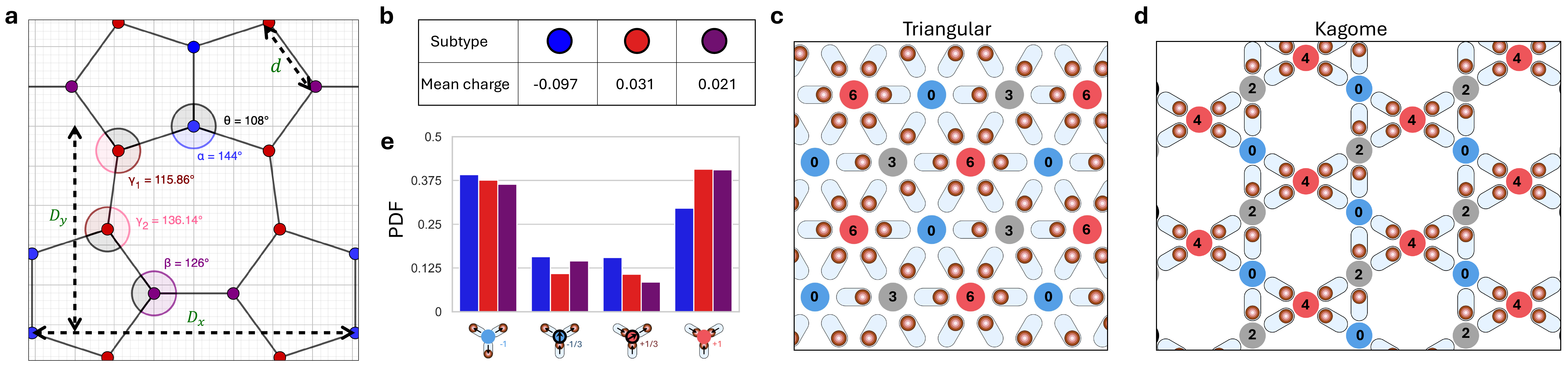}
\caption{\textbf{Construction of the Pentaheptite lattice.}  \textbf{a}
Schematic showing the angles and distances that result when constructing the pentaheptite lattice imposing an equal distance between the vertices. (\textbf{b}) Configurations of the four vertex types that results from the presence of the different angles when constructing the pentaheptite lattice. Top row shows the mean charge difference between the different configurations, and bottom the distribution of occurrence of these configurations in a lattice.(\textbf{c,d}) Schematic showing the low energy configuration predicted for the triangular (c) and the kagome (d) anti-spin ice in presence of strong attractive interactions.}
\label{figureSI1}
\end{figure*}

\begin{figure}[t]
\includegraphics[width=\columnwidth]{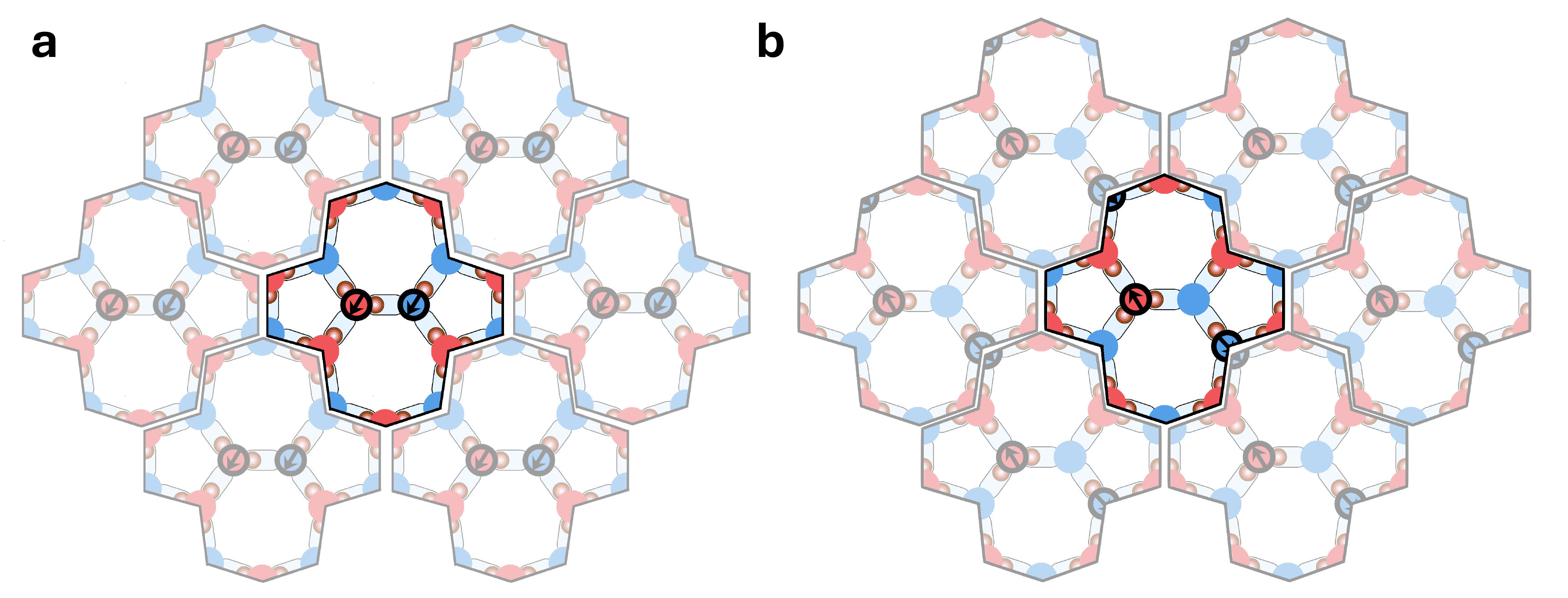}
\caption{\textbf{Pentaheptite unit cells.}  \textbf{a,b}
Two low energy configurations of the pentaheptite anti-spin ice obtained from repeating $6$ times the basic unit cell of the lattice made of $2$ pentagons and $2$ heptagons. 
The difference between both structures is that in (a) the $2$ frustrated vertices are adjacent, while in (b) they are separated by one vertex.  
The unit cell highlighted in the central region has  $8$ vertices : the $2$ vertices at center,  $8$ vertices shared between $2$ unit cells and  $6$ vertices shared by $3$ unit cells: $2+8/4+6/3=8$. Using this arrangement it is possible to compute the different statistical quantities measured at high field in the manuscript. The calculated spin-spin correlation value is $\langle Q_i \cdot Q_j \rangle = -128/216=-0.593$ for (a) and  $\langle Q_i \cdot Q_j \rangle = -126/216=-0.583$
for (b), very close to the values obtained in experiments and simulations, Fig.~\ref{figure3}(d).}
\label{figureSI2}
\end{figure}

\vspace{5px}
\noindent\textbf{Construction of the pentaheptite lattice.}  
We use the following recipe to construct an equilateral pentaheptite lattice, as also illustrated in Fig.~\ref{figureSI1}(a).
\begin{itemize}
\item First, we draw a regular pentagon with side length $d=1.1756$ if the pentagon is drawn in a circle of radius $R=1$.
\item After that, we draw a second regular pentagon that shares an edge with the first one. The common edge is vertical. It forms a "pentagon doublet".
\item We then draw a new pentagon doublet and link it to the first one
along the horizontal direction. The link has a side $d$, which results in a horizontal distance $D_x=4.08d$ between  the edges of two pentagons.We then repeat the process to get a "line" of pentagon doublets.
\item A second horizontal line of pentagon doublets is then drawn on top of it, shifted horizontally so that the vertical edges are aligned with the middle of the horizontal edges. We then link the two lines to close the heptagons. The vertical spacing is then adjusted so that the links have size $d$, resulting in a vertical distance $D_y=2.61d$, as shown in Fig.~\ref{figureSI1}(a).
\item We then repeat the operation several times to tessellate the plan with 
the repeating polygons.
\end{itemize}

As shown in Fig.~\ref{figureSI1}(a), a side effect of this construction is the emergence of different bond angles between the vertices: $\alpha=144\,^{\circ}$, $\beta=126\,^{\circ}$, $\gamma_1=115,86\,^{\circ}$, $\gamma_1=136,14\,^{\circ}$, and $\theta=108\,^{\circ}$.
Taking into account these angles introduce a further, small degeneracy in the four types of vertices. Each vertex type may be present in three different configurations that are characterized by a similar energetic value 
within $1/\%$ as shown in Fig.~\ref{figureSI1}(b).

Finally,   Figs.~\ref{figureSI1}(c,d) show the expected low energy state for two non-bipartite geometries in the presence of attractive interactions, the triangular ($z=6$) in  Fig.~\ref{figureSI1}(c) and the kagome ($z=4$) lattices in  Fig.~\ref{figureSI1}(c). 

\section*{Supplementary Videos}
In the article are $5$ video clips in support of the findings in the main text.
\begin{itemize}
\item \textbf{Supplementary Video 1 (.MOV)}: This video clip shows the collapse of two paramagnetic colloidal particles (diameter $\sigma = 2.8 \rm{\mu m}$) subjected 
to an in-plane rotating magnetic field with frequency $f= 50$ Hz and 
amplitude $B=4$mT. The fast spinning field induces time-averaged attractive interactions which lead to the formation of a spinning pair. The video is in real time and corresponds to 
the data shown in Figure 1(c) of the article.

\item \textbf{Supplementary Video 2 (.MOV)} Simulation of $N=840$ particles placed on a square lattice ($z=4$) and subjected to a rotating magnetic field with frequency $f=50$ Hz 
growing linearly from $B=0$ mT
to $B_{\rm{max}}=1.2$ mT at a rate  $\alpha = 1.2\cdot 10^{-3} \, \rm{mT s^{-1}}$. 
The video has been accelerated $30\times$ and 
corresponds to Figure 2(a) of the main text.

\item \textbf{Supplementary Video 3 (.MOV)} Simulation of $N=1279$ paramagnetic colloids on a honeycomb geometry ($z=3$) under a rotating  field ($f=50$ Hz, $\alpha = 1.2 \cdot 10^{-3} \, \rm{mT s^{-1}}$, $B_{\rm{max}}=1.2$ mT). 
The video has been accelerated $30\times$ and 
corresponds to Figure 2(b) of the main text.

\item \textbf{Supplementary Video 4 (.MOV)} Simulation of $N=1980$ paramagnetic colloids on a pentaheptite lattice ($z=3$). The applied field has frequency $f=50$Hz and grows linearly to  $B_{\rm{max}}=2$ mT at a rate $\alpha = 2 \cdot 10^{-3} \, \rm{mT s^{-1}}$. 
The video has been accelerated $30\times$ and 
corresponds to Figure 3(a) of the main text.

\item \textbf{Supplementary Video 5 (.MOV)} Experimental realization of the pentaheptite anti-spin ice with $N=205$ paramagnetic colloids. The applied field has a driving frequency $f=1$ Hz and its amplitude increases every $60$ s with an increment $\Delta B=0.375$ mT ending at $B_{\rm{max}}=3.75$ mT. The total duration time is $\Delta t = 600$ s and the field rate is given by $\alpha = 6.25\cdot 10^{-3} \, \rm{mT s^{-1}}$). 
The video has been accelerated $30\times$ and 
corresponds to Figure 3(b) of the main text.

\end{itemize}

\section*{Data availability}
The authors declare that all data supporting the findings of this
study are available in the paper and its Supplementary Information
files or available from the corresponding authors upon request.

\section*{Data availability}
The codes used in this study are available from the corresponding
author upon request.

\section*{Acknowledgments}

This work has received funding from the 
European Research Council (ERC) under the European Union's Horizon 2020 research and innovation programme (grant agreement no. 811234). 
The work of C.N. was performed under the auspices of the U.S. Department of Energy (DOE) at Los Alamos National Laboratory, operated by Triad National Security, LLC (contract 89233218CNA000001). The project was supported by the Laboratory Directed Research and Development (LDRD) program at LANL.
P. T. acknowledges support from the Ministerio
de Ciencia e Innovaci\'on (Project No. PID2022-137713NB-C22),
the Ag\`encia de Gesti\'o d'Ajuts Universitaris i de
Recerca (Project No. 2021 SGR 00450) and the Generalitat de Catalunya under Program ``ICREA Acad\`emia''.

\section*{Author Contributions}  

R. B. and M. T. performed the experiments. R. B. ran the numerical simulations.  C. N. developed the theoretical model. P. T. conceived the idea and supervised the work. The paper was written by C. N. and P. T. with input from all of the co-authors. All authors
discussed the results and commented on the manuscript at all stages.

\section*{Competing interests}  
The authors declare no competing interests.

\section*{Additional information}  
\textbf{Supplementary Information} is available in the online version
of the paper.

\vspace{5px}

\textbf{Correspondence} and requests for materials
related to the experiments should be addressed to P.T.\
(ptierno@ub.edu).

\end{document}